\documentclass[10pt,english]{article}
\usepackage[T1]{fontenc}
\usepackage[latin9]{inputenc}
\usepackage{geometry}
\usepackage{textcomp}
\usepackage{amsmath}
\usepackage{amssymb}
\usepackage{esint}

\makeatletter
\newenvironment{lyxcode}
{\par\begin{list}{}{
\setlength{\rightmargin}{\leftmargin}
\setlength{\listparindent}{0pt}
\raggedright
\setlength{\itemsep}{0pt}
\setlength{\parsep}{0pt}
\normalfont\ttfamily}%
 \item[]}
{\end{list}}

\@ifundefined{date}{}{\date{}}
\numberwithin{equation}{section}

\makeatother

\usepackage{babel}
\begin{document}

\title{
   \hfill {\small {USTC-ICTS-16-14}}\\
   Sound waves in the compactified D0-D4 brane system
}

\maketitle
\begin{center}
\footnote{Email: edlov@mail.ustc.edu.cn}Wenhe Cai\emph{$^{*,\dagger}$}
and \footnote{Email: cloudk@mail.ustc.edu.cn}Si-wen Li\emph{$^{*}$}
\par\end{center}

\begin{center}
\emph{$^{*}$Department of Modern Physics, }\\
\emph{ University of Science and Technology of China, }\\
\emph{ Hefei 230026, Anhui, China}\\
\emph{ }
\par\end{center}

\begin{center}
\emph{$^{\dagger}$Interdisciplinary Center for Theoretical Study,
}\\
\emph{ University of Science and Technology of China, }\\
\emph{ Hefei 230026, Anhui, China}
\par\end{center}

\vspace{16mm}

\begin{abstract}
As an extension to our previous work, we study the transport properties
of the Witten-Sakai-Sugimoto model in the black D4-brane background
with smeared D0-branes (D0-D4/D8 system). Because of the presence
of the D0-branes, in the bubble configuration this model is holographically
dual to 4d QCD or Yang-Mills theory with a Chern-Simons term. And
the number density of the D0-branes corresponds to the coupling constant
($\theta$ angle) of the Chern-Simons term in the dual field theory.
In this paper, we accordingly focus on small number density of the
D0-branes to study the sound mode in the black D0-D4 brane system
since the coupling of the Chern-Simons term should be quite weak in
QCD. Then we derive its 5d effective theory and analytically compute
the speed of sound and the sound wave attenuation in the approach
of Gauge/Gravity duality. Our result shows the speed of sound and
the sound wave attenuation is modified by the presence of the D0-branes.
Thus they depend on the $\theta$ angle or chiral potential in this
holographic description.
\end{abstract}
\newpage{}

\section{Introduction}

Since the Gauge/Gravity duality or AdS/CFT correspondence was proposed,
it has become a valuable tool in analyzing near-equilibrium dynamics
of the strongly coupled plasma for a long time \cite{key-01,key-02,key-03,key-04,key-05}.
Recent years, many people believe that the QCD quark-gluon plasma
(QGP) has been produced in heavy ion collision experiments at RHIC
\cite{key-06,key-07,key-08}, thus the potential application of Gauge/Gravity
or AdS/CFT correspondence for the hydrodynamic description of QGP
becomes the most important motivation of the research in this direction.
On the other hand, as one famous top-down prototype of holographic
QCD, the Witten-Sakai-Sugimoto model \cite{key-09,key-10,key-11}
introduces a supergravity description based on the geometric background
generated by $N_{c}$ D4-branes compactified on a cycle. Naturally,
studying the strongly coupled hydrodynamics of QCD has become one
of the most interesting aspects of this holographic model \cite{key-12,key-13,key-14}.

In the Witten-Sakai-Sugimoto model, it is realizing dual \cite{key-15,key-16}
to four-dimensional QCD-like theory in the large-$N_{c}$ limit. Specifically,
the D4-branes are compactified on a cycle with appropriate boundary
conditions, therefore the dual field theory is non-conformal and non-supersymmetric
which couples to the ``Kaluza-Klein'' field in the adjoint representation.
Additionally there are $N_{f}$ species of massless flavored quarks
introduced by embedding $N_{f}$ pairs of probe $\mathrm{D8}/\overline{\mathrm{D8}}$-branes.
In the D4 solitonic solution, the flavor $\mathrm{D8}/\overline{\mathrm{D8}}$-branes
are connected at the IR region, which holographically corresponds
to the broken chiral symmetry in the dual field theory. And the light
mesons come from the world volume theory on the connected $\mathrm{D8}/\overline{\mathrm{D8}}$-branes
in its low-energy effective theory.

Previously, the setup of the original Witten-Sakai-Sugimoto model
was parallel implanted into the D4-brane background with smeared D0-branes
\cite{key-17} as an extension (i.e. D0-D4/D8 brane system). And this
system is holographically dual to the QCD or Yang-Mills theory with
a topological term (i.e. the Chern-Simons term). It would be more
clear if we take into account the action of the D4-branes in the presence
of the smeared D0-branes,

\begin{equation}
S_{D_{4}}=-\mu_{4}\mathrm{Tr}\int d^{4}xdx^{4}e^{-\phi}\sqrt{-\det\left(\mathcal{G}+\mathcal{F}\right)}+\mu_{4}\int C_{5}+\frac{1}{2}\mu_{4}\int C_{1}\wedge\mathcal{F}\wedge\mathcal{F},\label{eq:1}
\end{equation}
where $\mu_{4}=\left(2\pi\right)^{-4}l_{s}^{-5}$, $l_{s}$ is the
size of the string, $\mathcal{G}$ is the induced metric and $\mathcal{F}=2\pi\alpha^{\prime}F$
which is proportional to the gauge field strength on the D4-brane.
$C_{1},\ C_{5}$ is the Romand-Romand 1- and 5- form respectively
and $x^{4}$ represents the periodic direction which is wrapped on
the cycle. Obviously, the Yang-Mill action comes from the leading
order of the first part in (\ref{eq:1}) (i.e. the Dirac-Born-Infield
(DBI) action) if it could be expanded by small $\mathcal{F}$. In
the bubble background of the Witten-Sakai-Sugimoto in the D0-D4 background,
we have the solution $C_{1}\sim\theta dx^{4}$ \cite{key-17}, thus
the last term in (\ref{eq:1}) could be integrated as,

\begin{equation}
\int_{S_{x^{4}}}C_{1}\sim\theta,\ \ \ \ \int_{S_{x^{4}}\times\mathbb{R}^{4}}C_{1}\wedge\mathcal{F}\wedge\mathcal{F}\sim\theta\int_{\mathbb{R}^{4}}\mathcal{F}\wedge\mathcal{F}.\label{eq:2}
\end{equation}
Consequently, the theory on the worldvolume of the D4-branes holographically
corresponds to the QCD or Yang-Mills theory with a topological term
as shown in (\ref{eq:2}). Phenomenologically, this topological term
(\ref{eq:2}) may lead to some observable effects such as the Chiral
Magnetic Effect (CME) (or some other effect in the glueball condensation)
\cite{key-18,key-19}. Therefore, while the experimental upper bounds
on $\theta$ is quite small, the $\theta$-dependence in QCD or Yang-Mills
theory is very interesting. Motivated by these, we have had many previous
works on this D0-D4/D8 holographic system as \cite{key-20,key-21,key-22}
(also see other people's work about the holographic $\theta$-dependence
in \cite{key-23,key-24}). Due to the relation between $\theta$ angle
and chiral potential \cite{key-25}, in this manuscript, we would
like to extend the study to the hydrodynamics in the black D0-D4 system
although the dual field theory of this system is less clear\footnote{While the confined geometry of the original Witten-Sakai-Sugimoto
model corresponds to the confinement phase, it is less clear for the
deconfined geometry (black brane background) in the dual field theory.
It has been discussed in \cite{key-26,key-27} and also in our previous
study \cite{key-28,key-29}. In this sense, as a implanted version
of the original Witten-Sakai-Sugimoto model, the dual field theory
of the black D0-D4/D8 system is also less clear.}.

The background geometry of the black D0-D4 brane system also satisfies
condition of \cite{key-30,key-31}, so the shear viscosity $\eta$
saturates the universal viscosity bound as in \cite{key-32},

\begin{equation}
\frac{\eta}{s}=\frac{1}{4\pi},\label{eq:3}
\end{equation}
where $s$ is the entropy density. It shows $\eta/s$ should not be
affected by the presence of the D0-branes (in other words, the $\theta$
angle). Hence we are going to take a next step toward understanding
transport phenomena in four-dimensional gauge plasma, i.e. to study
the sound waves in this holographic system. And many researches of
the sound waves by holographic duality could be reviewed, such as
in \cite{key-33,key-34}, also in the original Witten-Sakai-Sugimoto
model \cite{key-12}. Therefore as a generalization and comparison
to the present results in \cite{key-12}, it would be quite interesting
to consider the influence of the $\theta$ angle or chiral potential
on the sound mode by this holographic system.

In this paper, after this introduction part, we will review the geometry
of the black D0-D4 system briefly in section 2. Then we derive the
5-dimensionally effective theory of this system in section 3. Because
of the presence of the D0-branes, it shows there should be a vector
field in the effective theory which is additional to \cite{key-12}.
Interestingly, this vector might relate to some other observable effect\footnote{This additional vector may be related to the Chiral Vortical Separation
Effect (CVSE) \cite{key-14}, we would like to take a future study
on it in our framework.}. Moreover, we find our effective theory is also similar to the resultant
theory in \cite{key-35,key-36}. In section 4, we study the fluctuations
of the relevant fields in its effective theory and discuss how to
simplify the following computations for the sound mode. Then in section
5, the speed of the sound and the sound wave attenuation are accordingly
calculated in the hydrodynamic limit. We find they are affected by
the presence of the D0-branes and it may be interpreted as the modification
from the $\theta$ angle (\ref{eq:2}) in the viewpoint of the dual
field theory, or in other words, the speed of the sound and the sound
wave attenuation depend on the chiral potential ($\mu_{5}$) in this
holographic description. The final section is the summary and discussion
of this paper.

\section{Review of D0-D4 background}

We are going to review the black D0-D4 system briefly in this section,
and some results have been presented in \cite{key-17,key-20,key-22,key-21,key-37,key-38}.
In Einstein frame, the black brane solution of $N_{c}$ D4-brane with
$N_{0}$ smeared D0-branes reads \cite{key-17,key-38},

\begin{eqnarray}
ds^{2} & = & H_{4}^{-\frac{3}{8}}\left[-H_{0}^{-\frac{7}{8}}f_{T}\left(U\right)\left(dx^{0}\right)^{2}+H_{0}^{\frac{1}{8}}\delta_{ij}dx^{i}dx^{j}+H_{0}^{\frac{1}{8}}\left(dS^{1}\right)^{2}\right]+H_{4}^{\frac{5}{8}}H_{0}^{\frac{1}{8}}\left[\frac{dU^{2}}{f_{T}\left(U\right)}+U^{2}d\Omega{}_{4}^{2}\right]\nonumber \\
e^{-\left(\Phi-\Phi_{0}\right)} & = & H_{4}^{1/4}/H_{0}^{3/4},\ \ F_{2}=\frac{1}{\sqrt{2!}}\frac{\mathcal{A}}{U^{4}}\frac{1}{H_{0}^{2}}dU\wedge dx^{0},\ \ F_{4}=\frac{1}{\sqrt{4!}}\mathcal{B}\epsilon_{4}.\label{eq:4}
\end{eqnarray}
where

\begin{eqnarray}
\mathcal{A} & = & \frac{\left(2\pi l_{s}\right)^{7}g_{s}N_{0}}{\omega_{4}V_{4}},\ \ \mathcal{B}=\frac{\left(2\pi l_{s}\right)^{3}g_{s}N_{c}}{\omega_{4}},\ \ e^{\Phi_{0}}=g_{s},\nonumber \\
H_{4} & = & 1+\frac{U_{Q_{4}}^{3}}{U^{3}},\ \ H_{0}=1+\frac{U_{Q_{0}}^{3}}{U^{3}},\ \ f_{T}\left(U\right)=1-\frac{U_{\Lambda}^{3}}{U^{3}}.
\end{eqnarray}
we have used $g_{s},\ d\Omega_{4},\ \epsilon_{4},\ \omega_{4}=8\pi^{2}/3$
to represent the string coupling, the line element, the volume form
and the volume of a unit $S^{4}$ respectively. $U_{\Lambda}$ represents
the position of the horizon and $V_{4}$ is the volume of the D4-brane.
Notice that $x^{4}$ is the periodic direction and the D0-branes have
been smeared in the $x^{i},\ i=1,2,3$ and $x^{4}$ directions homogeneously.
Moreover, for the reader convenience, the relation between the integration
parameters $\mathcal{A},\ \mathcal{B},\ U_{Q_{0}}$ and $U_{Q_{4}}$
is given as \cite{key-17},

\begin{equation}
\mathcal{A}=3\sqrt{U_{Q_{0}}^{3}\left(U_{Q_{0}}^{3}+U_{\Lambda}^{3}\right)},\ \ \mathcal{B}=3\sqrt{U_{Q_{4}}^{3}\left(U_{Q_{4}}^{3}+U_{\Lambda}^{3}\right)}.
\end{equation}
By taking the near horizon limit i.e. fixing $U/\alpha^{\prime}$
and $U_{\Lambda}/\alpha^{\prime}$ where $\alpha^{\prime}=l_{s}^{2}$,
we have the following relations,

\begin{eqnarray}
U_{Q_{4}}^{3} & \rightarrow & \pi\alpha^{\prime3/2}g_{s}N_{c}=\frac{\beta g_{YM}^{2}N_{c}l_{s}^{2}}{4\pi}\equiv R^{3},\nonumber \\
H_{4}\left(U_{\Lambda}\right) & \rightarrow & \frac{R^{3}}{U_{\Lambda}^{3}},\nonumber \\
\beta & \rightarrow & \frac{4\pi}{3}U_{\Lambda}^{-1/2}R^{3/2}H_{0}^{1/2}\left(U_{\Lambda}\right).
\end{eqnarray}
where $\beta$ is the size of the periodic time direction. Consequently,
in the near horizon limit the black brane solution (\ref{eq:4}) becomes

\begin{eqnarray}
ds^{2} & = & \left(\frac{U}{R}\right)^{\frac{9}{8}}\left[-H_{0}^{-\frac{7}{8}}f_{T}\left(U\right)\left(dx^{0}\right)^{2}+H_{0}^{\frac{1}{8}}\delta_{ij}dx^{i}dx^{j}+H_{0}^{\frac{1}{8}}\left(dS^{1}\right)^{2}\right]+\left(\frac{R}{U}\right)^{\frac{15}{8}}H_{0}^{\frac{1}{8}}\left[\frac{dU^{2}}{f_{T}\left(U\right)}+U^{2}d\Omega{}_{4}^{2}\right]\nonumber \\
e^{\Phi} & = & g_{s}\left(\frac{U}{R}\right)^{3/4}H_{0}^{3/4}.\label{eq:8}
\end{eqnarray}
Then the deformed relations in the presence of D0-branes to the variables
in QCD are as follows,
\begin{equation}
R^{3}=\frac{\lambda l_{s}^{2}}{2M_{KK}};\ g_{s}=\frac{\lambda}{2\pi M_{KK}N_{c}l_{s}};\ U_{\Lambda}=\frac{2}{9}M_{KK}\lambda l_{s}^{2}H_{0}\left(U_{\Lambda}\right).\label{eq:9}
\end{equation}
where $\lambda=g_{YM}^{2}N_{c}$ is the 't Hooft coupling constant
and $M_{KK}$ is the mass scale.

\section{Dimensional reduction to 5d theory}

As we are going to study the hydrodynamics by $\mathrm{AdS_{5}/CFT_{4}}$
duality and the bulk fields in the D0-D4 background are described
by 10-dimensional type IIA supergravity, so in this section let us
first employ the standard Kaluza-Klein reduction on $S^{1}\times S^{4}$
as \cite{key-12,key-33,key-36}, to derive the 5-dimensionally effective
theory of this system. In the Einstein frame, the 10d action of type
IIA supergravity is given as,

\begin{equation}
S_{IIA}=\frac{1}{2k_{0}^{2}}\int d^{10}x\sqrt{-G}\left[\mathcal{R}^{(10)}-\frac{1}{2}\nabla_{M}\Phi\nabla^{M}\Phi-e^{\Phi/2}\left|F_{4}\right|^{2}-e^{3\Phi/2}\left|F_{2}\right|^{2}\right]\label{eq:10}
\end{equation}
where $\phi$ is the dilaton and $F_{4},\ F_{2}$ are the Romand-Romand
4- and 2- form field strength respectively. We have used ``$G$''
to represent the determinant of 10-dimensional metric $G_{MN}$, where
the indexes $\left(M,\ N\right)$ run from 0 to 9. And the following
ansatz of the metric would be helpful for the dimensional reduction,

\begin{eqnarray}
ds_{\left(10\right)}^{2} & = & G_{MN}dx^{M}dx^{N}\nonumber \\
 & = & e^{-\frac{10}{3}f}g_{ab}dx^{a}dx^{b}+e^{2f+8w}\left(dS^{1}\right)^{2}+e^{2f-2w}d\Omega_{4}^{2},\label{eq:11}
\end{eqnarray}
where $x^{a}$ represents $x^{a}=\left\{ x^{\mu},\ U\right\} ,\ \mu=0,1...3$.
Furthermore, we have assumed that the $S^{1}\times S^{4}$ dependence
could be trivially reduced, it means the fields $f$ and $w$ do not
depend on $S^{1}\times S^{4}$. By using the ansatz (\ref{eq:11}),
we could obtain some useful relations which are

\begin{align}
\sqrt{-G} & =\sqrt{-g}e^{-\frac{10}{3}f}g_{4}^{1/2},\nonumber \\
\sqrt{-G}\left|F_{4}\right|^{2} & =\mathcal{B}^{2}e^{-8\left(f-w\right)}\sqrt{-g}e^{-\frac{10}{3}f}g_{4}^{1/2},\nonumber \\
\sqrt{-G}\left|F_{2}\right|^{2} & =\sqrt{-g}F_{ab}F_{cd}g^{ac}g^{bd}e^{\frac{10}{3}f}g_{4}^{1/2},\nonumber \\
\sqrt{-G}\nabla_{M}\Phi\nabla^{M}\Phi & =\sqrt{-g}g_{4}^{1/2}\nabla_{a}\Phi\nabla^{a}\Phi\label{eq:12}
\end{align}
where $g_{4}$ represents the determinant of the metric on $S^{4}$
and $g_{ab}$ is the 5d metric. The relation between 10d $\mathcal{R}^{\left(10\right)}$and
5d $\mathcal{R}^{\left(5\right)}$curvature scalar is given as\footnote{We will not give the full relation in (\ref{eq:13}) since there would
be some additional total derivatives if imposing the full relation
of (\ref{eq:13}) to the action (\ref{eq:10}). Those terms have thus
been dropped off. So only the relevant terms are given in (\ref{eq:13}).}

\begin{equation}
\mathcal{R}^{\left(10\right)}=e^{\frac{10}{3}f}\left[\mathcal{R}^{\left(5\right)}-20g^{ab}\partial_{a}w\partial_{b}w-\frac{40}{3}g^{ab}\partial_{a}f\partial_{b}f\right]+12e^{-2\left(f-w\right)}.\label{eq:13}
\end{equation}
After inserting (\ref{eq:13}) into (\ref{eq:10}) and integrating
over $S^{1}\times S^{4}$, we obtain the 5d effective action which
takes the following form,

\begin{equation}
S_{5d}=\frac{\pi\mathcal{V}_{4}}{k_{0}^{2}}\int d^{5}x\sqrt{-g}\left[\mathcal{R}^{\left(5\right)}-\frac{1}{2}g^{ab}\partial_{a}\Phi\partial_{b}\Phi-20g^{ab}\partial_{a}w\partial_{b}w-\frac{40}{3}g^{ab}\partial_{a}f\partial_{b}f-\mathcal{P}-e^{\frac{10}{3}f+\frac{3}{2}\Phi}F_{ab}F_{cd}g^{ac}g^{bd}\right],\label{eq:14}
\end{equation}
where $\mathcal{V}_{4}$ represents the volume of the 4-sphere and

\begin{equation}
\mathcal{P}=\mathcal{B}^{2}e^{\frac{\Phi}{2}-\frac{34}{3}f+8w}-12e^{-\frac{16}{3}f+2w}.
\end{equation}
The equations of motion for $\Phi,\ w,\ f$ and $g_{ab}$ could be
obtained from (\ref{eq:14}) which are as follows,

\begin{align}
g^{ab}\nabla_{a}\nabla_{b}f-\frac{3}{80}\frac{\partial\mathcal{P}}{\partial f}-\frac{1}{8}e^{\frac{10}{3}f+\frac{3}{2}\Phi}F_{ab}F_{cd}g^{ac}g^{bd} & =0,\nonumber \\
g^{ab}\nabla_{a}\nabla_{b}w-\frac{1}{40}\frac{\partial\mathcal{P}}{\partial w} & =0,\nonumber \\
g^{ab}\nabla_{a}\nabla_{b}\Phi-\frac{\partial\mathcal{P}}{\partial\Phi}-\frac{3}{2}e^{\frac{10}{3}f+\frac{3}{2}\Phi}F_{ab}F_{cd}g^{ac}g^{bd} & =0,\nonumber \\
\partial_{a}\left[\sqrt{-g}e^{\frac{10}{3}f+\frac{3}{2}\Phi}F^{ab}\right] & =0,\nonumber \\
\frac{1}{2}\partial_{a}\Phi\partial_{b}\Phi+20\partial_{a}w\partial_{b}w+\frac{40}{3}\partial_{a}f\partial_{b}f+\frac{1}{3}g_{ab}\mathcal{P}+\left(2F_{ca}F_{\ b}^{c}-\frac{1}{3}g_{ab}F_{cd}F^{cd}\right)e^{\frac{10}{3}f+\frac{3}{2}\Phi} & =\mathcal{R}_{ab}^{\left(5d\right)}.\label{eq:16}
\end{align}
We have used $F_{ab}$ to represent the components of the Romand-Romand
2-form $F_{2}$. Then let us consider the 5d ansatz of the metric
as

\begin{equation}
ds_{\left(5\right)}^{2}=-c_{1}^{2}dt^{2}+c_{2}^{2}\delta_{ij}dx^{i}dx^{j}+c_{3}^{2}dU^{2},
\end{equation}
which is obtained from the following corresponding 10d metric (\ref{eq:11}),

\begin{equation}
ds_{\left(10\right)}^{2}=e^{-\frac{10}{3}f}\left[-c_{1}^{2}dt^{2}+c_{2}^{2}\delta_{ij}dx^{i}dx^{j}+c_{3}^{2}dU^{2}\right]+e^{2f+8w}\left(dS^{1}\right)^{2}+e^{2\left(f-w\right)}d\Omega_{4}^{2}.\label{eq:18}
\end{equation}
Comparing (\ref{eq:18}) with the black brane solution of the Witten-Sakai-Sugimoto
model in the D0-D4 background (\ref{eq:4}), it leads to the following
relations,

\begin{align}
f & =\ \frac{1}{16}\log H_{0}+\frac{13}{80}\log U,\nonumber \\
w & =\ \frac{1}{10}\log U,\nonumber \\
c_{1} & =\ f_{T}^{1/2}U^{5/6}H_{0}^{-1/3},\nonumber \\
c_{2} & =\ H_{0}^{1/6}U^{5/6},\nonumber \\
c_{3} & =\ f_{T}^{-1/2}H_{0}^{1/6}U^{-2/3}.\label{eq:19}
\end{align}
where we have set $g_{s}=R=1$ for convenience (as a comparison with
\cite{key-12}), so that $\mathcal{B}=\sqrt{\frac{9}{2}}$. And one
can verify the reduced functions in (\ref{eq:19}) satisfies the 5d
effective equations of motion (\ref{eq:16}) consistently. Consequently
we obtain the 5d effective action (\ref{eq:14}) and its solution
(\ref{eq:18}) (\ref{eq:19}) of our D0-D4 brane system. But as a
difference from the original D4-brane system, there is an additional
vector field $C_{a}$ in the low energy effective theory whose field
strength is the Romand-Romand 2-form defined as $F_{ab}=\partial_{a}C_{b}-\partial_{b}C_{a}$.

\section{Fluctuations }

In this section, let us study the fluctuations of the relevant fields
in the black D0-D4 background by replacing\footnote{By the solution for the D0-D4 background (\ref{eq:4}), we have assumed
that only one component of $C_{a}$ is nonzero which is $C_{t}$. },

\begin{align}
g_{ab} & \rightarrow g_{ab}+h_{ab},\nonumber \\
f & \rightarrow f+\delta f,\nonumber \\
w & \rightarrow w+\delta w,\nonumber \\
\Phi & \rightarrow\Phi+\delta\Phi,\nonumber \\
C_{a} & \rightarrow C_{a}+\delta C_{a}.\label{eq:20}
\end{align}
where $\left\{ h_{ab},\ \delta f,\ \delta w,\ \delta\Phi,\ \delta C_{a}\right\} $
are the fluctuations while $\left\{ g_{ab},\ f,\ w,\ \Phi,\ C_{a}\right\} $
are the background configurations of the D0-D4 system i.e. the classical
solution of the equations of motion (\ref{eq:16}). For the fluctuations
of the metric, we are going to choose the following gauge as \cite{key-12,key-36,key-34,key-33},

\begin{equation}
h_{aU}=0.
\end{equation}
Furthermore, we have assumed that the fluctuations of the metric depends
on $\left\{ t,\ z,\ U\right\} $\footnote{The coordinate $x^{\mu}$ could be identified as $\left\{ t,\ x,\ y,\ z\right\} $.}
only i.e. the system we are considering is $O\left(2\right)$ rotationally
symmetric in the $x-y$ plane.

In the linearized case, the following sets of the metric are decoupled
from each other because of the $O\left(2\right)$ symmetry \cite{key-34},

\begin{align}
\left\{ h_{12}\right\}  & ,\nonumber \\
\left\{ h_{11}-h_{22}\right\}  & ,\nonumber \\
\left\{ h_{01},\ h_{13}\right\}  & ,\nonumber \\
\left\{ h_{02},\ h_{23}\right\}  & ,\nonumber \\
\left\{ h_{00},\ h_{\alpha\alpha}=h_{11}+h_{22},\ h_{03},\ h_{33}\right\}  & .\label{eq:22}
\end{align}
While the first three sets in (\ref{eq:22}) are related to the shear
modes, the last set corresponds to the sound waves which is the concern
in this manuscript. Besides, there are additional fluctuations as
$\left\{ \delta f,\ \delta w,\ \delta\Phi,\ \delta C_{a}\right\} $
from the dimension-reductional scalars and vector. As a comparison,
let us employ the similar conventions as \cite{key-12,key-33,key-36}
by introducing\footnote{It would not be confused with (\ref{eq:1}) (\ref{eq:2}) if we use
the same $\mathcal{F}$ to represent the fluctuation of the function
$f$ here.},

\begin{align}
h_{00}=e^{-i\omega t+iqx_{3}}c_{1}^{2}H_{tt},\  & h_{03}=e^{-i\omega t+iqx_{3}}c_{2}^{2}H_{tz},\nonumber \\
h_{\alpha\alpha}=e^{-i\omega t+iqx_{3}}c_{2}^{2}H_{\alpha\alpha},\  & h_{33}=e^{-i\omega t+iqx_{3}}c_{2}^{2}H_{zz},\nonumber \\
\delta f=e^{-i\omega t+iqx_{3}}\mathcal{F},\  & \delta w=e^{-i\omega t+iqx_{3}}\mathcal{W},\nonumber \\
\delta\Phi=e^{-i\omega t+iqx_{3}}\varphi,\  & \delta C_{a}=e^{-i\omega t+iqx_{3}}\mathcal{C}_{a}.\label{eq:23}
\end{align}
where the definition of the functions $c_{1}$ and $c_{2}$ has been
given in (\ref{eq:19}) and $\left\{ H_{tt},\ H_{tz},\ H_{\alpha\alpha},\ H_{zz},\ \mathcal{F},\ \mathcal{W},\ \varphi,\ \mathcal{C}_{a}\right\} $
are the functions which depend on the radial coordinate $U$ only.
By inserting (\ref{eq:20}) (\ref{eq:23}) into (\ref{eq:16}) and
expanding all the equations of motion at a linearized level, we obtain
five ordinary differential equations as in the Appendix. And the relevant
equations are collected as follows once we evaluate (\ref{eq:A-1})
- (\ref{eq:A-5}) by (\ref{eq:18}) (\ref{eq:19}) and (\ref{eq:23}),
which are

\begin{align}
0= & H_{tt}^{\prime\prime}+\left[\ln\frac{c_{1}^{2}c_{2}^{3}}{c_{3}}\right]^{\prime}H_{tt}^{\prime}-\left[\ln c_{1}\right]^{\prime}H_{ii}^{\prime}-\frac{c_{3}^{2}}{c_{1}^{2}}\left(q^{2}\frac{c_{1}^{2}}{c_{2}^{2}}H_{tt}+\omega^{2}H_{ii}+2\omega qH_{tz}\right)\nonumber \\
 & -\frac{2}{3}c_{3}^{2}\left(\frac{\partial\mathcal{P}}{\partial f}\mathcal{F}+\frac{\partial\mathcal{P}}{\partial w}\mathcal{W}+\frac{\partial\mathcal{P}}{\partial\Phi}\varphi\right)+\left(\frac{6c_{1}^{\prime}c_{2}^{\prime}}{c_{1}c_{2}}-\frac{2c_{1}^{\prime}c_{3}^{\prime}}{c_{1}c_{3}}+\frac{2c_{1}^{\prime\prime}}{c_{1}}+\frac{2}{3}c_{3}^{2}\mathcal{P}\right)H_{tt}\nonumber \\
 & +\frac{4F_{tU}}{9c_{1}^{2}}e^{\frac{10}{3}f+\frac{3}{2}\Phi}\left(12\hat{F}_{tU}+20F_{tU}\mathcal{F}+9F_{tU}\varphi\right),\label{eq:24}
\end{align}

\begin{align}
0= & H_{tz}^{\prime\prime}+\left[\ln\frac{c_{2}^{5}}{c_{1}c_{3}}\right]^{\prime}H_{tz}^{\prime}+q\omega\frac{c_{3}^{2}}{c_{2}^{2}}H_{\alpha\alpha}+\left(\frac{2c_{1}^{\prime}c_{2}^{\prime}}{c_{1}c_{2}}+\frac{4c_{2}^{\prime2}}{c_{2}^{2}}-\frac{2c_{2}^{\prime}c_{3}^{\prime}}{c_{2}c_{3}}+\frac{2c_{2}^{\prime\prime}}{c_{2}^{2}}+\frac{2}{3}c_{3}^{2}\mathcal{P}\right)H_{tz}\nonumber \\
 & \frac{4F_{tU}}{3c_{1}^{2}}e^{\frac{10}{3}f+\frac{3}{2}\Phi}\left(3\hat{F}_{3U}\frac{c_{1}^{2}}{c_{2}^{2}}+F_{tU}H_{tz}\right),\label{eq:25}
\end{align}

\begin{align}
0= & H_{aa}^{\prime\prime}+\left[\ln\frac{c_{1}c_{2}^{5}}{c_{3}}\right]^{\prime}H_{aa}^{\prime}+\frac{c_{3}^{2}}{c_{1}^{2}}\left(\omega^{2}-q^{2}\frac{c_{1}^{2}}{c_{2}^{2}}\right)H_{aa}+\left(H_{zz}^{\prime}-H_{tt}^{\prime}\right)\left[\ln c_{2}^{2}\right]^{\prime}\nonumber \\
 & +\left(\frac{4c_{1}c_{3}c_{2}^{\prime2}+2c_{2}c_{3}c_{1}^{\prime}c_{2}^{\prime}-2c_{2}c_{1}c_{2}^{\prime}c_{3}^{\prime}}{c_{1}c_{2}^{2}c_{3}}+\frac{2c_{2}^{\prime\prime}}{c_{2}}+\frac{2}{3}c_{3}^{2}\mathcal{P}\right)H_{aa}+\frac{4}{3}c_{3}^{2}\left(\frac{\partial\mathcal{P}}{\partial f}\mathcal{F}+\frac{\partial\mathcal{P}}{\partial w}\mathcal{W}+\frac{\partial\mathcal{P}}{\partial\Phi}\varphi\right)\nonumber \\
 & +\frac{4F_{tU}}{9c_{1}^{2}}e^{\frac{10}{3}f+\frac{3}{2}\Phi}\left(20F_{tU}\mathcal{F}+12\hat{F}_{tU}+6F_{tU}H_{tt}+3F_{tU}H_{aa}+9F_{tU}\varphi\right),\label{eq:26}
\end{align}

\begin{align}
0= & H_{zz}^{\prime\prime}+\left[\ln\frac{c_{1}c_{2}^{4}}{c_{3}}\right]^{\prime}H_{zz}^{\prime}+\left(H_{aa}^{\prime}-H_{tt}^{\prime}\right)\left[\ln c_{2}\right]^{\prime}+\frac{c_{3}^{2}}{c_{1}^{2}}\left[\omega^{2}H_{zz}+2\omega qH_{tz}+q^{2}\frac{c_{1}^{2}}{c_{2}^{2}}\left(H_{tt}-H_{aa}\right)\right]\nonumber \\
 & +H_{zz}\left(\frac{c_{1}^{\prime}c_{2}^{\prime}}{c_{1}c_{2}}+\frac{c_{2}^{\prime2}}{c_{2}^{2}}-\frac{c_{2}^{\prime}c_{3}^{\prime}}{c_{2}c_{3}}+\frac{2c_{2}^{\prime\prime}}{c_{2}}+\frac{2}{3}c_{3}^{2}\mathcal{P}\right)+\frac{2}{3}c_{3}^{2}\left(\frac{\partial\mathcal{P}}{\partial f}\mathcal{F}+\frac{\partial\mathcal{P}}{\partial w}\mathcal{W}+\frac{\partial\mathcal{P}}{\partial\Phi}\varphi\right)\nonumber \\
 & +\frac{2F_{tU}}{9c_{1}^{2}}e^{\frac{10}{3}f+\frac{3}{2}\Phi}\left(12\hat{F}_{tU}+20F_{tU}\mathcal{F}+6F_{tU}H_{tt}+6F_{tU}H_{zz}+9F_{tU}\varphi\right),\label{eq:27}
\end{align}

\begin{align}
0= & \mathcal{F}^{\prime\prime}+\left[\ln\frac{c_{1}c_{2}^{3}}{c_{3}}\right]^{\prime}\mathcal{F}^{\prime}+\frac{1}{2}f^{\prime}\left(H_{ii}^{\prime}-H_{tt}^{\prime}\right)+\frac{c_{3}^{2}}{c_{1}^{2}}\left(\omega^{2}-q^{2}\frac{c_{1}^{2}}{c_{2}^{2}}\right)\mathcal{F}-\frac{3}{80}c_{3}^{2}\left(\frac{\partial^{2}\mathcal{P}}{\partial f^{2}}\mathcal{F}+\frac{\partial^{2}\mathcal{P}}{\partial f\partial w}\mathcal{W}+\frac{\partial^{2}\mathcal{P}}{\partial f\partial\Phi}\varphi\right)\nonumber \\
 & +\frac{F_{tU}}{24c_{1}^{2}}e^{\frac{10}{3}f+\frac{3}{2}\Phi}\left(12\hat{F}_{tU}+20F_{tU}\mathcal{F}+6F_{tU}H_{tt}+9F_{tU}\varphi\right),\label{eq:28}
\end{align}

\begin{align}
0= & \mathcal{W}^{\prime\prime}+\left[\ln\frac{c_{1}c_{2}^{3}}{c_{3}}\right]^{\prime}\mathcal{W}^{\prime}+\frac{1}{2}w^{\prime}\left(H_{ii}^{\prime}-H_{tt}^{\prime}\right)+\frac{c_{3}^{2}}{c_{1}^{2}}\left(\omega^{2}-q^{2}\frac{c_{1}^{2}}{c_{2}^{2}}\right)\mathcal{W}\nonumber \\
 & -\frac{1}{40}c_{3}^{2}\left(\frac{\partial^{2}\mathcal{P}}{\partial w\partial f}\mathcal{F}+\frac{\partial^{2}\mathcal{P}}{\partial w^{2}}\mathcal{W}+\frac{\partial^{2}\mathcal{P}}{\partial w\partial\Phi}\varphi\right),\label{eq:29}
\end{align}

\begin{align}
0= & \varphi^{\prime\prime}+\left[\ln\frac{c_{1}c_{2}^{3}}{c_{3}}\right]^{\prime}\varphi^{\prime}+\frac{1}{2}\Phi^{\prime}\left(H_{ii}^{\prime}-H_{tt}^{\prime}\right)+\frac{c_{3}^{2}}{c_{1}^{2}}\left(\omega^{2}-q^{2}\frac{c_{1}^{2}}{c_{2}^{2}}\right)\varphi-c_{3}^{2}\left(\frac{\partial^{2}\mathcal{P}}{\partial\Phi\partial f}\mathcal{F}+\frac{\partial^{2}\mathcal{P}}{\partial\Phi\partial w}\mathcal{W}+\frac{\partial^{2}\mathcal{P}}{\partial\Phi^{2}}\varphi\right)\nonumber \\
 & +\frac{F_{tU}}{2c_{1}^{2}}e^{\frac{10}{3}f+\frac{3}{2}\Phi}\left(12\hat{F}_{tU}+20F_{tU}\mathcal{F}+6F_{tU}H_{tt}+9F_{tU}\varphi\right),\label{eq:30}
\end{align}
where $\hat{F}_{ab}=\mathcal{A}e^{i\omega t-iqx_{3}}\left(\partial_{a}\delta C_{b}-\partial_{b}\delta C_{a}\right)$\footnote{By this definition, there would be an ``$i$'' factor in $\hat{F}_{tz}$
once we calculate the derivative respected to $z$ or $x_{3}$. In
this sense, (\ref{eq:32}) is a real equation although there is an
``$i$'' factor.}. Besides, there are three additional first order constraints which
comes by associating with the (partially) fixed diffeomorphism invariance\footnote{In fact there are more additional equations from the vanished components
of linearized Ricci tensor. We have checked that those equations determine
the vanished components of $\delta F_{ab}$. As a result, the nonzero
and relevant components of $\delta F_{ab}$ mixed to the sound mode
are only $\delta F_{tz},\ \delta F_{zU},\ \delta F_{tU}$. So the
nonzero components of $\delta C_{a}$ could be $\delta C_{t}$ only
if $C_{U},\ \delta C_{U}$ are gauged by $C_{U},\ \delta C_{U}=0$. },

\begin{align}
0= & \omega\left(H_{ii}^{\prime}+\left[\ln\frac{c_{2}}{c_{1}}\right]^{\prime}H_{ii}\right)+q\left(H_{tz}^{\prime}+2\left[\ln\frac{c_{2}}{c_{1}}\right]^{\prime}H_{tz}\right)\nonumber \\
 & +\omega\left(\frac{80}{3}f^{\prime}\mathcal{F}+40w^{\prime}\mathcal{W}+\Phi^{\prime}\varphi\right),\label{eq:31}
\end{align}

\begin{align}
0= & q\left(H_{tt}^{\prime}-\left[\ln\frac{c_{2}}{c_{1}}\right]^{\prime}H_{tt}\right)+\frac{c_{2}^{2}}{c_{1}^{2}}\omega H_{tz}^{\prime}-qH_{aa}^{\prime}+4i\hat{F}_{tz}\frac{F_{tU}}{c_{1}^{2}}e^{\frac{10}{3}f+\frac{3}{2}\Phi}\nonumber \\
 & -q\left(\frac{80}{3}f^{\prime}\mathcal{F}+40w^{\prime}\mathcal{W}+\Phi^{\prime}\varphi\right),\label{eq:32}
\end{align}

\begin{align}
0= & \left[\ln c_{1}c_{2}^{2}\right]^{\prime}H_{ii}^{\prime}-\left[\ln c_{2}^{3}\right]^{\prime}H_{tt}^{\prime}+\frac{c_{3}^{2}}{c_{1}^{2}}\left[\omega^{2}H_{ii}+2\omega qH_{tz}+q^{2}\frac{c_{1}^{2}}{c_{2}^{2}}\left(H_{tt}-H_{aa}\right)\right]\nonumber \\
 & +c_{3}^{2}\left(\frac{\partial\mathcal{P}}{\partial f}\mathcal{F}+\frac{\partial\mathcal{P}}{\partial w}\mathcal{W}+\frac{\partial\mathcal{P}}{\partial\Phi}\varphi\right)-\left(\frac{80}{3}f^{\prime}\mathcal{F}^{\prime}+40w^{\prime}\mathcal{W}^{\prime}+\Phi^{\prime}\varphi^{\prime}\right)+\frac{2}{3}c_{3}^{2}\mathcal{P}\left(H_{ii}-H_{tt}\right)\nonumber \\
 & +\left(\frac{c_{1}^{\prime}c_{2}^{\prime}}{c_{1}c_{2}}+\frac{2c_{2}^{\prime2}}{c_{2}^{2}}-\frac{c_{2}^{\prime}c_{3}^{\prime}}{c_{2}c_{3}}+\frac{c_{2}^{\prime\prime}}{c_{2}}\right)H_{ii}+\left(\frac{c_{1}^{\prime}c_{3}^{\prime}}{c_{1}c_{3}}-\frac{3c_{1}^{\prime}c_{2}^{\prime}}{c_{1}c_{2}}-\frac{c_{1}^{\prime\prime}}{c_{1}}\right)H_{tt}\nonumber \\
 & +\frac{F_{tU}}{3c_{1}^{2}}e^{\frac{10}{3}f+\frac{3}{2}\Phi}\left(12\hat{F}_{tU}+20F_{tU}\mathcal{F}+10F_{tU}H_{tt}+2F_{tU}H_{aa}+2F_{tU}H_{zz}+9F_{tU}\varphi\right).\label{eq:33}
\end{align}
Notice that, we do not give the relations of the fluctuations from
the equation of motion for the vector field $C_{a}$, since this vector
part corresponds to diffusive or transverse channel \cite{key-34},
which are less relevant to the sound modes \cite{key-34}. Therefore
we will not attempt to discuss more about the vector part thus we
can simply set $\delta C_{a}=0$ if studying the sound mode only.
Nevertheless, the surviving equations of motion from (\ref{eq:24})
- (\ref{eq:33}) are still complicated. On the other hand, since the
parameter $\mathcal{A}$ is related the coupling constant of the topological
term in the dual field theory as (\ref{eq:2}), which is actually
very small, thus it simplifies the calculation greatly if we only
consider the leading order in small $\mathcal{A}$ expansion of all
the equations in (\ref{eq:24}) - (\ref{eq:33}). Then if we introduce
the gauge invariant variables as \cite{key-12,key-33,key-34,key-36},

\begin{align}
Z_{H} & =4\frac{q}{\omega}H_{tz}+2H_{zz}-H_{aa}\left(1-\frac{q^{2}}{\omega^{2}}\frac{c_{1}^{\prime}c_{1}}{c_{2}^{\prime}c_{2}}\right)+2\frac{q^{2}}{\omega^{2}}\frac{c_{1}^{2}}{c_{2}^{2}}H_{tt},\nonumber \\
Z_{f} & =\mathcal{F}-\frac{f^{\prime}}{\left[\ln c_{2}^{4}\right]^{\prime}}H_{aa},\nonumber \\
Z_{w} & =\mathcal{W}-\frac{w^{\prime}}{\left[\ln c_{2}^{4}\right]^{\prime}}H_{aa},\nonumber \\
Z_{\Phi} & =\varphi-\frac{\Phi^{\prime}}{\left[\ln c_{2}^{4}\right]^{\prime}}H_{aa}.
\end{align}
with a new coordinate

\begin{equation}
x=\frac{c_{1}}{c_{2}},
\end{equation}
then we find the decoupled equations of motion for $Z$' s by imposing
(\ref{eq:24}) - (\ref{eq:33}) in small $\mathcal{A}$ expansion
as

\begin{align}
0 & =\frac{d^{2}Z_{H}}{dx^{2}}+\left[\frac{3q^{2}\left(2x^{2}-1\right)+5\omega^{2}}{x\left(5\omega^{2}-q^{2}\left(3+2x^{2}\right)\right)}+\mathcal{A}^{2}h_{1}\left(x\right)\right]\frac{dZ_{H}}{dx}\nonumber \\
 & +\left[\frac{4}{9}\frac{\left(-\omega^{2}+q^{2}x^{2}\right)\left(q^{2}\left(3+2x^{2}\right)-5\omega^{2}\right)-18q^{2}U_{\Lambda}x^{2}\left(1-x^{2}\right)^{5/3}}{\left(5\omega^{2}-q^{2}\left(3+2x^{2}\right)\right)\left(1-x^{2}\right)^{5/3}x^{2}U_{\Lambda}}+\mathcal{A}^{2}h_{2}\left(x\right)\right]Z_{H}\nonumber \\
 & +\left[\frac{4}{15}\frac{q^{2}\left(-3q^{2}+5\omega^{2}\right)}{\omega^{2}\left(5\omega^{2}-q^{2}\left(3+2x^{2}\right)\right)}+\mathcal{A}^{2}g_{1}\left(x\right)\right]\kappa+\mathcal{O}\left(\mathcal{A}^{4}\right),\nonumber \\
0 & =\frac{d^{2}\kappa}{dx^{2}}+\left[\frac{1}{x}+\mathcal{A}^{2}g_{2}\left(x\right)\right]\frac{d\kappa}{dx}+\left[\frac{4\left(\omega^{2}-q^{2}x^{2}\right)}{9U_{\Lambda}x^{2}\left(1-x^{2}\right)^{5/3}}+\mathcal{A}^{2}g_{3}\left(x\right)\right]\kappa+\mathcal{O}\left(\mathcal{A}^{4}\right).\label{eq:36}
\end{align}
where

\begin{align}
\kappa & =48Z_{w}+9Z_{\phi}+52Z_{f},\nonumber \\
h_{1}\left(x\right) & =\frac{2\left(q^{4}\left(54-33x^{2}+84x^{4}+20x^{6}\right)-10q^{2}\left(9+16x^{4}\right)\omega^{2}+125x^{2}\omega^{4}\right)}{45xU_{\Lambda}^{6}\left(q^{2}\left(3+2x^{2}\right)-5\omega^{2}\right)^{2}},\nonumber \\
h_{2}\left(x\right) & =\frac{4}{1215x^{2}\left(-1+x^{2}\right)^{2}U_{\Lambda}^{7}\left(q^{2}\left(3+2x^{2}\right)-5\omega^{2}\right)}\times\nonumber \\
 & \ \ \bigg[5q^{6}x^{2}\left(1-x^{2}\right)^{1/3}\left(-27+36x^{4}+16x^{6}\right)+125\left(3-4x^{2}\right)\left(1-x^{2}\right)^{1/3}\omega^{6}\nonumber \\
 & \ \ +15q^{2}\omega^{2}\left(27\left(-1+x^{2}\right)^{3}\left(-3+2x^{2}\right)U_{\Lambda}+5\left(1-x^{2}\right)^{1/3}\left(-6-x^{2}+12x^{4}\right)\omega^{2}\right)\nonumber \\
 & \ \ -3q^{4}\left(27\left(-1+x^{2}\right)^{3}\left(-9+4x^{2}\right)U_{\Lambda}+5\left(1-x^{2}\right)^{1/3}\left(-9-30x^{2}+32x^{4}+32x^{6}\right)\omega^{2}\right)\bigg].
\end{align}
We will not give the explicit formula of the functions $g_{1,2,3}\left(x\right)$
used in (\ref{eq:36}) since they are too messy and lengthy. Moreover,
in the next section, it would be clear that the functions $g_{1,2,3}\left(x\right)$
are actually less useful to the calculations for the sound mode, because
the sound mode are relevant to $Z_{H}$ only.

\section{Hydrodynamic limit in small $\mathcal{A}$ expansion}

In this section, let us study the physical fluctuation equations (\ref{eq:36})
in the hydrodynamic limit i.e. $\omega\rightarrow0,\ q\rightarrow0$
but $\frac{\omega}{q}$ is fixed as a constant. As many discussions,
only the leading and next-to-leading solution (in small $q$ expansion)
of (\ref{eq:36}) is needed. On the hand, since the sound mode are
relevant to $Z_{H}$ instead of $\kappa$, so similar to the discussions
and calculations in \cite{key-12,key-33,key-34,key-36}, we can simply
choose $\kappa=0$ as the solution for (\ref{eq:36}) consistently\footnote{It has been discussed that $\kappa=0$ could be a solution for $\mathcal{A}=0$
case \cite{key-12}, so it is also consistent with this solution for
$\kappa$ in small $\mathcal{A}$ case}. And for $Z_{H}$ we find that at the horizon $x\rightarrow0_{+}$,
$Z_{H}\rightarrow x^{\pm\frac{i\omega}{2\pi T}}x^{-\mathcal{A},\ }$\footnote{As $\mathcal{A}>0$, it means $x^{-\mathcal{A}}$ is also singular
if $x\rightarrow0$. This behavior at the horizon is a bit different
from the original D4-brane system, however it should be consistent
in the small $\mathcal{A}$ limit.}. By imposing the incoming boundary condition on all physical modes,
we assume that

\begin{equation}
Z_{H}=x^{-\frac{i\omega}{2\pi T}}x^{-\mathcal{A}}z_{H},\label{eq:38}
\end{equation}
where $z_{H}$ must be regular at horizon. Additionally, since we
are interested in the hydrodynamic pole dispersion-relation in the
stress-energy correlation, it would be convenient to parameterize
the $\boldsymbol{\omega}$ and $\boldsymbol{\mathfrak{q}}$ as

\begin{equation}
\mathfrak{\boldsymbol{\omega}}=v_{s}\boldsymbol{\mathfrak{q}}-i\boldsymbol{\mathfrak{q}}^{2}\Gamma.\label{eq:39}
\end{equation}
where
\begin{lyxcode}
\begin{equation}
\mathfrak{\boldsymbol{\omega}}=\frac{\omega}{2\pi T},\ \boldsymbol{\mathfrak{q}}=\frac{q}{2\pi T},
\end{equation}

\end{lyxcode}
and $v_{s},\ \Gamma$ is the speed of sound, the sound wave attenuation
respectively, which would be determined from the pole dispersion relation.
Without the loss of generality, we can choose the boundary condition
for $z_{H}$  as \cite{key-12},

\begin{equation}
z_{H}\bigg|_{x\rightarrow0_{+}}=1,\ z_{H}\bigg|_{x\rightarrow1_{-}}=0.\label{eq:41}
\end{equation}
By expanding $z_{H}$ with small $\boldsymbol{\mathfrak{q}}$, we
assume

\begin{equation}
z_{H}=z_{H,0}+i\boldsymbol{\mathfrak{q}}z_{H,1}.
\end{equation}
 Inserting (\ref{eq:38}) - (\ref{eq:41}) into (\ref{eq:36}) in
small $\boldsymbol{\mathfrak{q}}$ and $\mathcal{A}$ expansion with
$\kappa=0$, we obtain the following equations for $z_{H,0}$ and
$z_{H,1}$ as

\begin{align}
0 & =z_{H,0}^{\prime\prime}-\frac{6x^{2}+5v_{s}^{2}-3}{x\left(2x^{2}-5v_{s}^{2}+3\right)}z_{0,H}^{\prime}+\frac{8}{2x^{2}-5v_{s}^{2}+3}z_{0,H}\nonumber \\
 & \ \ +\mathcal{A}\left[\left(\frac{1}{x^{2}}+\frac{6x^{2}+5v_{s}^{2}-3}{x^{2}\left(2x^{2}-5v_{s}^{2}+3\right)}\right)z_{0,H}-\frac{2}{x}z_{0,H}^{\prime}\right]\label{eq:43}
\end{align}
for leading order in $\mathcal{O}\left(\boldsymbol{\mathfrak{q}}^{0}\right)$
and

\begin{align}
0 & =z_{H,1}^{\prime\prime}+\frac{3-5v_{s}^{2}-6x^{2}}{x\left(3-5v_{s}^{2}+2x^{2}\right)}z_{H,1}^{\prime}+\frac{8}{3-5v_{s}^{2}+2x^{2}}z_{H,1}\nonumber \\
 & \ \ +\frac{2v_{s}\left(40x^{2}\Gamma+20x^{2}v_{s}^{2}-25v_{s}^{4}+30v_{s}^{2}-4x^{4}-12x^{2}-9\right)}{x\left(2x^{2}-5v_{s}^{2}+3\right)^{2}}z_{H,0}^{\prime}\nonumber \\
 & \ \ -\frac{8v_{s}\left(-2x^{2}+5v_{s}^{2}-3+10\Gamma\right)}{\left(2x^{2}-5v_{s}^{2}+3\right)^{2}}z_{H,0}-\mathcal{A}\bigg[\frac{2z_{H,1}^{\prime}}{x}-\frac{8}{3-5v_{s}^{2}+2x^{2}}z_{H,1}\nonumber \\
 & \ \ +\frac{2v_{s}\left(40x^{2}\Gamma+20x^{2}v_{s}^{2}-25v_{s}^{4}+30v_{s}^{2}-4x^{4}-12x^{2}-9\right)}{x^{2}\left(2x^{2}-5v_{s}^{2}+3\right)^{2}}z_{H,0}\bigg].\label{eq:44}
\end{align}
for next to leading order in $\mathcal{O}\left(\boldsymbol{\mathfrak{q}}^{1}\right)$.
Then the solution for equation (\ref{eq:43}) can be obtained as,

\begin{align}
z_{H,0}= & \frac{\left(3-3\mathcal{A}-5v_{s}^{2}+5\mathcal{A}v_{s}^{2}-2x^{2}-2\mathcal{A}x^{2}\right)C_{1}}{1-5\mathcal{A}-5v_{s}^{2}+5\mathcal{A}v_{s}^{2}}+\frac{\left(3-3\mathcal{A}-5v_{s}^{2}+5\mathcal{A}v_{s}^{2}-2x^{2}-2\mathcal{A}x^{2}\right)}{2\left(1+\mathcal{A}\right)^{2}\left(1-5\mathcal{A}-5v_{s}^{2}+5\mathcal{A}v_{s}^{2}\right)}\nonumber \\
 & \times x^{\mathcal{A}}\left[\frac{1}{\mathcal{A}}-\frac{4\left(-3+5v_{s}^{2}\right)}{\left(-1+\mathcal{A}\right)\left(-3+5v_{s}^{2}+2x^{2}+\mathcal{A}\left(3-5v_{s}^{2}+2x^{2}\right)\right)}\right]C_{2}.\label{eq:45}
\end{align}
where $C_{1,2}$ are two integration constants. We impose the boundary
condition (\ref{eq:41}) for $z_{H,0}$,

\begin{equation}
z_{H,0}\bigg|_{x\rightarrow1^{-}}=0,
\end{equation}
Besides, in order to compare our solution (\ref{eq:45}) with \cite{key-12},
we further require

\begin{equation}
z_{H,0}\bigg|_{x\rightarrow0^{-}}=1.
\end{equation}
Thus the relation between the integration constants $C_{1,2}$ is
obtained as,

\begin{equation}
C_{2}=-\frac{2\mathcal{A}\left(-1+\mathcal{A}^{2}\right)^{2}C_{1}^{2}}{4\mathcal{A}+C_{1}-2\mathcal{A}C_{1}+\mathcal{A}^{2}C_{1}}.\label{eq:48}
\end{equation}
The solution (\ref{eq:45}) should be definitely able to return to
\cite{key-12} in small $\mathcal{A}$ limit, in this sense we have
the following extra relations

\begin{equation}
C_{1}=\frac{1-5v_{s}^{2}}{3-5v_{s}^{2}},\ \ C_{2}\sim\mathcal{O}\left(\mathcal{A}^{4}\right)\label{eq:49}
\end{equation}
Accordingly, it yields,

\begin{align}
v_{s} & \simeq\frac{1}{\sqrt{5}}-\frac{2\mathcal{A}}{\sqrt{5}}+\mathcal{O}\left(\mathcal{A}^{2}\right).\label{eq:50}
\end{align}
So the speed of the sound wave is shifted shown as (\ref{eq:50}).
Then the equation of motion for $z_{H,1}$ could be derived from (\ref{eq:44})
by inserting the solution for $z_{H,0}$ (\ref{eq:45}) (\ref{eq:50})
(\ref{eq:48}) and (\ref{eq:49}) similarly. However the resultant
equation from (\ref{eq:44}) is too complicated to be solved analytically.
Hence we expand $z_{H,1}$ in the small $\mathcal{A}$ case as,

\begin{equation}
z_{H,1}\left(x\right)=\mathcal{X}\left(x\right)+\mathcal{\mathcal{A}}\mathcal{Y}\left(x\right)+\mathcal{O}\left(\mathcal{A}^{2}\right).\label{eq:51}
\end{equation}
By the expansion (\ref{eq:51}), we obtain the decoupled equation
for $\mathcal{X}\left(x\right)$ in leading order $\mathcal{O}\left(\mathcal{A}^{0}\right)$
is,

\begin{equation}
0=\mathcal{X}^{\prime\prime}+\frac{\left(1-3x^{2}\right)}{x\left(1+x^{2}\right)}\mathcal{X}^{\prime}+\frac{4}{1+x^{2}}\mathcal{X}+\frac{8-20\Gamma}{\sqrt{5}\left(1+x^{2}\right)}\label{eq:52}
\end{equation}
and the equation for $\mathcal{Y}\left(x\right)$ in the next to the
leading order $\mathcal{O}\left(\mathcal{A}^{1}\right)$ is,

\begin{align}
0= & \mathcal{Y}^{\prime\prime}+\frac{-2\left(-1+x^{2}\right)^{2}\mathcal{X}^{\prime}+\left(1-2x^{2}-3x^{4}\right)\mathcal{Y}^{\prime}}{x\left(1+x^{2}\right)^{2}}+\frac{4\left(-1+x^{2}\right)^{2}\mathcal{X}+4\left(1+x^{2}\right)\mathcal{Y}}{\left(1+x^{2}\right)^{2}}\nonumber \\
 & +\frac{4-4x^{6}+x^{4}\left(-22+125\Gamma\right)+x^{2}\left(-46+205\Gamma\right)}{2\sqrt{5}x^{2}\left(1+x^{2}\right)^{2}}\label{eq:53}
\end{align}
The solution for (\ref{eq:52}) (\ref{eq:53}) could be found as

\begin{align}
\mathcal{X}\left(x\right)= & \frac{1}{\sqrt{5}}\left(5\Gamma-2\right)+D_{1}\left(-1+x^{2}\right)+D_{2}\left(-2-\ln x+x^{2}\ln x\right)\nonumber \\
\mathcal{Y}\left(x\right)= & \left(-1+x^{2}\right)E_{1}+E_{2}\left(-2-\ln x+x^{2}\ln x\right)+\frac{1}{40}\big(2\sqrt{5}-165\sqrt{5}\Gamma\nonumber \\
 & -160D_{2}-160D_{2}\ln x-32\sqrt{5}\ln x-8\sqrt{5}\ln^{2}x+8\sqrt{5}x^{2}\ln^{2}x-40D_{2}\ln^{2}x+40x^{2}D_{2}\ln^{2}x\big)\label{eq:54}
\end{align}
where $D_{1,2}$, $E_{1,2}$ are the integration constants. Since
we have required that $z_{H,1}$ must be regular at the horizon, it
yields

\begin{equation}
E_{2}=\frac{1}{\sqrt{5}\mathcal{A}},\ D_{2}=-\frac{1}{\sqrt{5}}.\label{eq:55}
\end{equation}
Then imposing the boundary condition (\ref{eq:41}) to (\ref{eq:51})
with (\ref{eq:54}), we obtain

\begin{align}
D_{1} & =-\mathcal{A}E_{1},\nonumber \\
\Gamma & \simeq\frac{2}{5}+\frac{4}{5}\mathcal{A}+\mathcal{O}\left(\mathcal{A}^{2}\right),\label{eq:56}
\end{align}
which shows how the sound wave attenuation is also shifted by $\mathcal{A}$.

\section{Summary and discussion}

In this paper, first we have studied the 5d effective theory of our
black D0-D4 system by the dimensional reduction (Kaluza-Klein reduction).
It contains an additional vector field to the 5d effective theory
of the original D4-brane system as \cite{key-12}. Then employing
the similar technique as \cite{key-12,key-33}, the physical fluctuations
in the 5d effective theory of our D0-D4 system has also been studied.
Although the dual field theory of the black D0-D4 system is not completely
clear (also for the original D4-brane system), we calculate the speed
of sound and the sound wave attenuation in the hydrodynamic limit
by using our 5d effective theory. Particularly, we focus on the small
$\mathcal{A}$ expansion in our calculations since the parameter $\mathcal{A}$
is related to the $\theta$ angle (the number density of D0-branes)
which should be very small in QCD.

While our solution of the gauge invariant variable (\ref{eq:45})
(\ref{eq:51}) is quite different, it could be able to return to \cite{key-12}
if expanded by small $\mathcal{A}$. Accordingly, it is allowed to
compare our result with \cite{key-12}. (\ref{eq:50}) and (\ref{eq:56})
show the speed of sound and the sound wave attenuation are all shifted
by the presence of the D0-branes and they return to \cite{key-12}
consistently if setting $\mathcal{A}=0$ (i.e. no D0-branes). Due
to (\ref{eq:2}), (\ref{eq:50}) and (\ref{eq:56}) could be interpreted
as the modification from the $\theta$ angle or the chiral potential
to the speed of sound and the sound wave attenuation. In hydrodynamics,
the speed of sound depends on the mass of Fermions and Bosons, also
the temperature \cite{key-33}. But our holographic result suggests
an additional $\theta$ dependence or chiral potential dependence
if the topological term of QCD or Yang-Mills theory is considered.
And as the leading order modification from the topological term (the
$\theta$ angle) , it shows the speed of sound decreases while the
sound wave attenuation increases\footnote{We have just noted a recent work \cite{key-39} which studies the
same holographic system after submitting the first version of this
manuscript to arXiv. And our result is qualitatively similar as \cite{key-39}}.

Besides, we need to keep in mind the simplification and the approximation
used in our calculations. First we do not consider the fluctuations
from the vector since this part is less relevant to the sound mode
\cite{key-13,key-34}, so we simply turn off this part. However it
may lead to some observable effects such as CVSE in hydrodynamics.
Thus a future study about it would be interesting and natural. Second,
the solution for $\kappa$ (the combination of the gauge invariant
variable with $Z_{w},\ Z_{f},\ Z_{\Phi}$) has been chosen as $\kappa=0$.
This solution for $\kappa$ is a rough choice while it is consistent
with its equation of motion (\ref{eq:36}) (also consistent with its
boundary condition in \cite{key-12}). Therefore a further improvement
to take into account the solution of $\kappa$ is also needed although
it might not change the results about the sound modes qualitatively.

\section*{Acknowledgments}

We would like to thank Chao Wu and Yi Yang for valuable comments and
discussions. We thank the Tsinghua University, Beijing, for its kind
hospitality during the program \textquotedblleft String 2016\textquotedblright ,
where this work was finalized.

\section*{Appendix}

In this appendix, we collect the equations of motion for the fluctuations
in the 5d effective theory. From (\ref{eq:16}) at the linearized
level, the equations for the fluctuations are\index{Commands!T!tag@\textbackslash{}tag}

\begin{align}
g^{ab}\left[\partial_{a}\partial_{b}\delta f-\left(\Gamma_{\ ab}^{c}\partial_{c}\delta f+\Gamma_{\ \ ab}^{(1)c}\partial_{c}f\right)\right]-h^{ab}\nabla_{a}\nabla_{b}f-\frac{3}{80}\left(\frac{\partial^{2}\mathcal{P}}{\partial f\partial\Phi}\delta\Phi+\frac{\partial^{2}\mathcal{P}}{\partial f^{2}}\delta f+\frac{\partial^{2}\mathcal{P}}{\partial f\partial w}\delta w\right)\nonumber \\
-\frac{1}{8}e^{\frac{10}{3}f+\frac{3}{2}\Phi}\left(2F^{ab}\delta F_{ab}-h^{ac}g^{bd}F_{ab}F_{cd}-F_{ab}F_{cd}g^{ac}h^{bd}+\frac{10}{3}F_{ab}F^{ab}\delta f+\frac{3}{2}F_{ab}F^{ab}\delta\Phi\right) & =0,\tag{A1}\label{eq:A-1}
\end{align}

\begin{equation}
g^{ab}\left[\partial_{a}\partial_{b}\delta w-\left(\Gamma_{\ ab}^{c}\partial_{c}\delta w+\Gamma_{\ \ ab}^{(1)c}\partial_{c}w\right)\right]-h^{ab}\nabla_{a}\nabla_{b}w-\frac{1}{40}\left(\frac{\partial^{2}\mathcal{P}}{\partial w\partial\Phi}\delta\Phi+\frac{\partial^{2}\mathcal{P}}{\partial w^{2}}\delta w+\frac{\partial^{2}\mathcal{P}}{\partial w\partial f}\delta f\right)=0,\tag{A2}\label{eq:A-2}
\end{equation}

\begin{align}
g^{ab}\left[\partial_{a}\partial_{b}\delta\Phi-\left(\Gamma_{\ ab}^{c}\partial_{c}\delta\Phi+\Gamma_{\ \ ab}^{(1)c}\partial_{c}\Phi\right)\right]-h^{ab}\nabla_{a}\nabla_{b}\Phi-\left(\frac{\partial^{2}\mathcal{P}}{\partial\Phi^{2}}\delta\Phi+\frac{\partial^{2}\mathcal{P}}{\partial\Phi\partial f}\delta f+\frac{\partial^{2}\mathcal{P}}{\partial\Phi\partial w}\delta w\right)\nonumber \\
-\frac{3}{2}e^{\frac{10}{3}f+\frac{3}{2}\Phi}\left(2F^{ab}\delta F_{ab}-h^{ac}g^{bd}F_{ab}F_{cd}-F_{ab}F_{cd}g^{ac}h^{bd}+\frac{10}{3}F_{ab}F^{ab}\delta f+\frac{3}{2}F_{ab}F^{ab}\delta\Phi\right) & =0,\tag{A3}\label{eq:A-3}
\end{align}

\begin{align}
\partial_{a}\left[\sqrt{-g}e^{\frac{10}{3}f+\frac{3}{2}\Phi}\left(\frac{10}{3}F^{ab}\delta f+\frac{3}{2}F^{ab}\delta\Phi+g^{ac}g^{bd}\delta F_{cd}-F_{cd}h^{ca}g^{bd}-F_{cd}g^{ca}h^{bd}\right)\right] & =0,\tag{A4}\label{eq:A-4}
\end{align}

\begin{align}
\frac{40}{3}\left(\partial_{a}f\partial_{b}\delta f+\partial_{a}\delta f\partial_{b}f\right)+20\left(\partial_{a}w\partial_{b}\delta w+\partial_{a}\delta w\partial_{b}w\right)+\frac{1}{2}\left(\partial_{a}\Phi\partial_{b}\delta\Phi+\partial_{a}\delta\Phi\partial_{b}\Phi\right)\nonumber \\
+\frac{1}{3}g_{ab}\left(\frac{\partial\mathcal{P}}{\partial f}\delta f+\frac{\partial\mathcal{P}}{\partial w}\delta w+\frac{\partial\mathcal{P}}{\partial\Phi}\delta\Phi\right)+\frac{1}{3}h_{ab}\mathcal{P}+\left(\frac{10}{3}\delta f+\frac{3}{2}\delta\Phi\right)\left(2F_{ca}F_{\ b}^{c}-\frac{1}{3}g_{ab}F_{cd}F^{cd}\right)e^{\frac{10}{3}f+\frac{3}{2}\Phi}\nonumber \\
+\left(2\delta F_{ca}F_{\ b}^{c}+2F_{\ a}^{c}\delta F_{cb}-2F_{ca}F_{db}h^{cd}-\frac{2}{3}g_{ab}F^{cd}\delta F_{cd}+\frac{2}{3}g_{ab}F_{\ c}^{d}F_{de}h^{ce}-\frac{1}{3}h_{ab}F_{cd}F^{cd}\right)e^{\frac{10}{3}f+\frac{3}{2}\Phi} & =\mathcal{R}_{ab}^{(1)}.\tag{A5}\label{eq:A-5}
\end{align}
where $\Gamma_{\ ab}^{c}$ and $\Gamma_{\ \ ab}^{(1)c}$ are defined
as,

\begin{align}
h^{ab}= & g^{ac}g^{bd}h_{cd},\nonumber \\
\Gamma_{\ ab}^{c}= & \frac{1}{2}g^{dc}\left(\partial_{b}g_{da}+\partial_{a}g_{db}-\partial_{d}g_{ab}\right),\nonumber \\
\Gamma_{\ \ ab}^{(1)c}= & \frac{1}{2}\left[g^{cd}\left(\partial_{a}h_{db}+\partial_{b}h_{ad}-\partial_{d}h_{ab}\right)-h^{cd}\left(\partial_{a}g_{db}+\partial_{b}g_{ad}-\partial_{d}g_{ab}\right)\right].\tag{A6}
\end{align}
and $\mathcal{R}_{ab}^{(1)}$ is defined as,

\begin{align}
\mathcal{R}_{ab}^{(1)}= & \partial_{a}\Gamma_{\ \ cb}^{(1)c}-\partial_{b}\Gamma_{\ \ ca}^{(1)c}+\Gamma_{\ \ ad}^{(1)c}\Gamma_{\ cb}^{d}+\Gamma_{\ \ ad}^{c}\Gamma_{\ \ cb}^{(1)d}\nonumber \\
 & -\Gamma_{\ \ bd}^{(1)c}\Gamma_{\ ca}^{d}-\Gamma_{\ \ bd}^{c}\Gamma_{\ \ ca}^{(1)d}.\tag{A7}
\end{align}
The equations (\ref{eq:24}) - (\ref{eq:33}) could be obtained from
(\ref{eq:A-1}) - (\ref{eq:A-5}).

\end{document}